\documentclass[amsmath,amssymb,
 aps]{revtex4}
\usepackage{graphicx,epsfig,pstricks,rotating}
\usepackage{amssymb}
\usepackage{amsfonts}
\usepackage{amsmath}
\usepackage[T2A]{fontenc}

\usepackage{graphics}
\usepackage{epsfig}
\begin{document}
\thispagestyle{empty}

\title{Pion damping width and pion spectral function in  hot pion gas}

\author{Goderidze D.}
\affiliation{Joint Institute for Nuclear Research, Joliot-Curie, 6,
Dubna, 141980, Russia}

\author{Friesen A.V.}
\email{avfriesen@theor.jinr.ru} \affiliation{Joint Institute for Nuclear Research, Joliot-Curie, 6, Dubna, 141980, Russia}

\author{Kalinovskiy Yu.L.}
\email{kalinov@jinr.ru}
\affiliation{Joint Institute for Nuclear Research, Joliot-Curie, 6\\
Dubna, 141980, Russia}


\begin{abstract}
The temperature behaviour of the pion width in the hadronic phase is investigated in the framework of the NJL model. The contribution to the width from the pion-pion collision  is considered with a scalar sigma-meson as an intermediate state. It is shown that the pion width  significantly broadens at $T>0.1$ GeV. Using the two-step iteration method, suggested by Kadanoff and Baym, the pion spectral function in a hot pion gas is calculated at different temperatures.
\begin{description}
\item[PACS numbers] 12.38.Mh; 25.75.Nq
\end{description}
\end{abstract}

\maketitle


\section{Introduction}
It is supposed that in ultra-relativistic heavy-ion collisions a new phase with free quarks and gluons (quark-gluon plasma) can be generated. This phase transition is accompanied by the restoration of spontaneously broken chiral symmetry. Theoretically, a spontaneous breaking of  symmetry leads to the existence of a massless Goldstone boson, which is supposed to be a pion in quantum chromodynamics (QCD). Being the lightest hadron with the Goldstone nature, the pion plays a special role in hadronic physics.  Elastic $\pi\pi$-scattering is a fundamental process for quantum chromodynamics (QCD) at low energies as it provides a direct link between the theoretical formalism of a chiral symmetry and experiment. Moreover, pions can be quickly created in the early phase of a heavy-ion collision, as two and more pions are the final state of some hadronic interactions. So the pion gas evolution at a finite temperature and density especially near the phase transition is an object of great interest. 

The increasing temperature of the pion gas leads to an increase in density ($n\sim T^3$) and in-medium collisions occur more often. It can make  the average lifetime of a particular pion state shorter and thus increase its width. With a further increase in  density, the raised inverse processes can restore the destroyed state and increase back the average lifetime of the pion state (or decrease its width). The resulting width is related to the returning process of a disturbed system to an equilibrium state (a damping width).  The occurrence of the finite pion damping width in dense matter was predicted by Blaschke et al. \cite{Voskresensky:1995tx}. Later, the effect of the finite pion width on the in-medium $\rho$-meson behaviour was discussed in Ref.~\cite{vanHees:2000bp}. 

In this article, we concentrate on the calculations of the pion damping width and the pion spectral function in the hot interacting pion gas below the critical temperature in the framework of the Nambu-Jona-Lasinio (NJL) model. This model is often used for investigation of in-medium meson properties  \cite{Blaschke:2003zt,Yudichev:2005yz}. 
The width of a particular state is calculated via the collision integral in the Kadanoff - Baym approach \cite{KadanoffBaym}, which includes  the total scattering amplitude of collision processes. The description of the $\pi\pi$-scattering amplitude in NJL-like models at the one-loop level includes two types of diagrams: the ''box" diagram and the meson exchange diagram \cite{Quack:1994vc,Fu:2009zs}, and is performed only in simplest kinematics (s=$4m_\pi^2, t=u=0$), as the calculations at finite temperature require the Matsubara technique, which is laborious in the arbitrary kinematics, especially for four-pole integrals\cite{Rehberg:1995nr,Khvorostukhin:2020foa}. We show that the $\pi\pi-$scattering amplitude for arbitrary kinematics can be truncated by using the pole approximation of meson propagators and supposing the 4-pion interaction to be a constant. The obtained relations are similar to the scattering amplitudes in the meson-exchange model \cite{Cotanch:2002vj}.

The spectral function of meson correlations can be a key to many real-time observables in strongly-interacting systems, as it provides information on quasi-particle spectra and collective excitation of the system. Speaking of  pion gas, some stages should be discussed: from the low-temperature non-interacting gas through the high-temperature non-relativistic interacted gas to the stage with high density, where the quark structure of the pion becomes more significant. The first stage can be described via a simple $\delta-$function. The second one is discussed, for example,  at $T<T_c$ in the framework of the O(4) linear $\sigma-$model \cite{Chiku:1997va}. In our work, the pion spectral function is calculated in the hot pion gas using the two-iteration method \cite{KadanoffBaym}, and this approach of the elastic scattering is also limited by  the temperature below the critical temperature $T < T_c\sim 0.19$ GeV where pion is still a bound state.  For the last stage, where $T>T_c$, the formalism of the phase shift of quark-antiquark scattering  to the meson spectral function can be applied in the framework of the NJL model\cite{Xia:2014bla}. The non-relativistic approach on the basis of  the Beth-Uhlenbeck formula, extended recently by D. Blaschke\cite{Blaschke:2019col,Blaschke:2013zaa} enables the authors to study  meson spectral functions at high temperatures where the chiral phase transition and the Mott transition have already appeared. These calculations show that there is still a chance for collective modes at $T>T_c$.

We will discuss the formalism of the NJL model and scattering amplitudes within the model in Section 2, and show numerical calculations
of the pion damping width and the pion spectral function in Section 3. Finally, we summarize in Section 4.

\section{Pion-pion scattering amplitude within the NJL model}
\label{sec:damping}
\subsection{The model formalism}

Generally, to consider pion-pion scattering in the medium, the $\sigma$- and $\rho$- mesons should be taken into account, as they appear as an intermediate state in the scattering process \cite{Quack:1994vc}. Nevertheless, the work focuses on the SU(2) NJL model \cite{RevModPhys.64.649,Kalinovsky:2015kzf}  and the Lagrangian with scalar and pseudo-scalar interactions has the following form:   
\begin{equation}
  \mathcal{L}_{\rm NJL}=\bar{q}\left(i\gamma_\mu \partial^\mu - \hat{m}_0
  \right) q+ G  \left[\left(\bar{q}q\right)^2+\left(\bar{q}i\gamma_5
      \vec{\tau} q \right)^2\right],
\label{njl}
\end{equation}
where $G$ is the scalar coupling constant, $\bar{q}, q$ are the quark
fields, $\hat{m}_0$ is the diagonal matrix of the current quark mass,
$\hat{m}_0 = {\rm diag}(m_u^0, m_d^0)$ with $m_u^0 = m_d^0 = m_0$, and
$\overrightarrow{\tau}$ are the Pauli matrices in space SU(2),
$\tau^a(a = 1,2,3)$.

In the mean field approximation, the constituent quark mass is
provided by the gap equation:
\begin{equation} 
m = m_0 + 2 i  G \int \frac{dp}{(2\pi)^4} {\rm Tr}\{S(p)\}, 
\end{equation}
where $S(p) = (\hat{p}-m)^{-1}$ is the quark propagator and trace is
taken over Dirac, flavour and colour indices.

Considering mesons as a quark-antiquark bound state, the meson propagator in the framework of the random-phase approximation can be written as a matrix in the meson space:
\begin{equation}
D_M(k^2) = \frac{2  G}{1 - 2 G \ \Pi_{M}(k^2)},
\label{mesonProp}
\end{equation}
where $k$ is the meson four-momentum. Equations for meson masses can be derived as poles of the meson propagator (Eq.(\ref{mesonProp})) at vanishing three-momentum (the Bethe-Salpeter equation):
\begin{equation}
\det\{1 - 2G \ \Pi_{M}(M_M-i\Gamma_M/2, \vec{0})\} = 0.
\label{Bet-Salp}
\end{equation}
The equation is given in complex form to remind that after the meson mass exceeds the mass of constituents, the equation does not represent the stable bound state, rather a resonant state. The complex form of Eq. (\ref{Bet-Salp}) is used to determine both mass $M_M$ and width $\Gamma_M$ of the meson. The polarization operator $\Pi_{M}(k^2)$ defines the meson
properties:
\begin{equation}
\Pi_{M} (k^2) = i \int \frac{d^4p}{(2\pi)^4} \ \mbox{Tr}\,
\left[ \Gamma_M S(p+k) \Gamma_M S(p)
 \right],   
\end{equation}
where the vertex factor $\Gamma_M$ depends on the sort of meson: $\Gamma_M = i \gamma_5 \tau^a $ for the pseudo-scalar meson and $\Gamma_M = {\bf 1} \tau^a $ for the scalar meson.  Both pion-quark  and $\sigma$-quark coupling strengths $g_{\pi qq}$, $g_{\sigma qq}$ are obtained from the  polarization operator $\Pi_M$ as
 \begin{equation}
     g^{-2}_{Mqq} = \frac{\partial \Pi_m (k^2)}{\partial k^2}\vert_{^{k^2 = M^2}}.
 \end{equation}

\begin{figure}[!h]
  \centerline{ \includegraphics[width=0.45\linewidth]{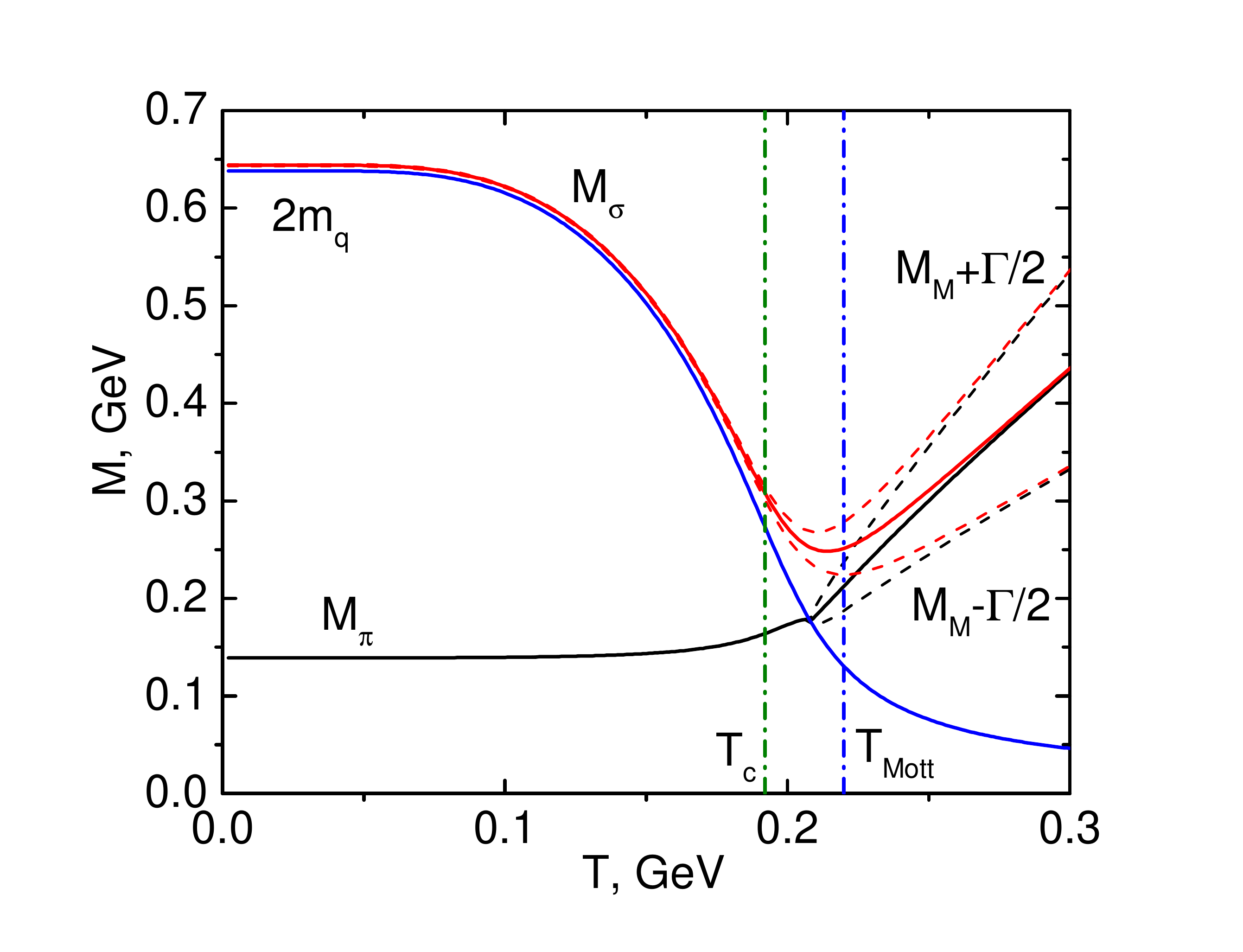}
  \includegraphics[width=0.45\linewidth]{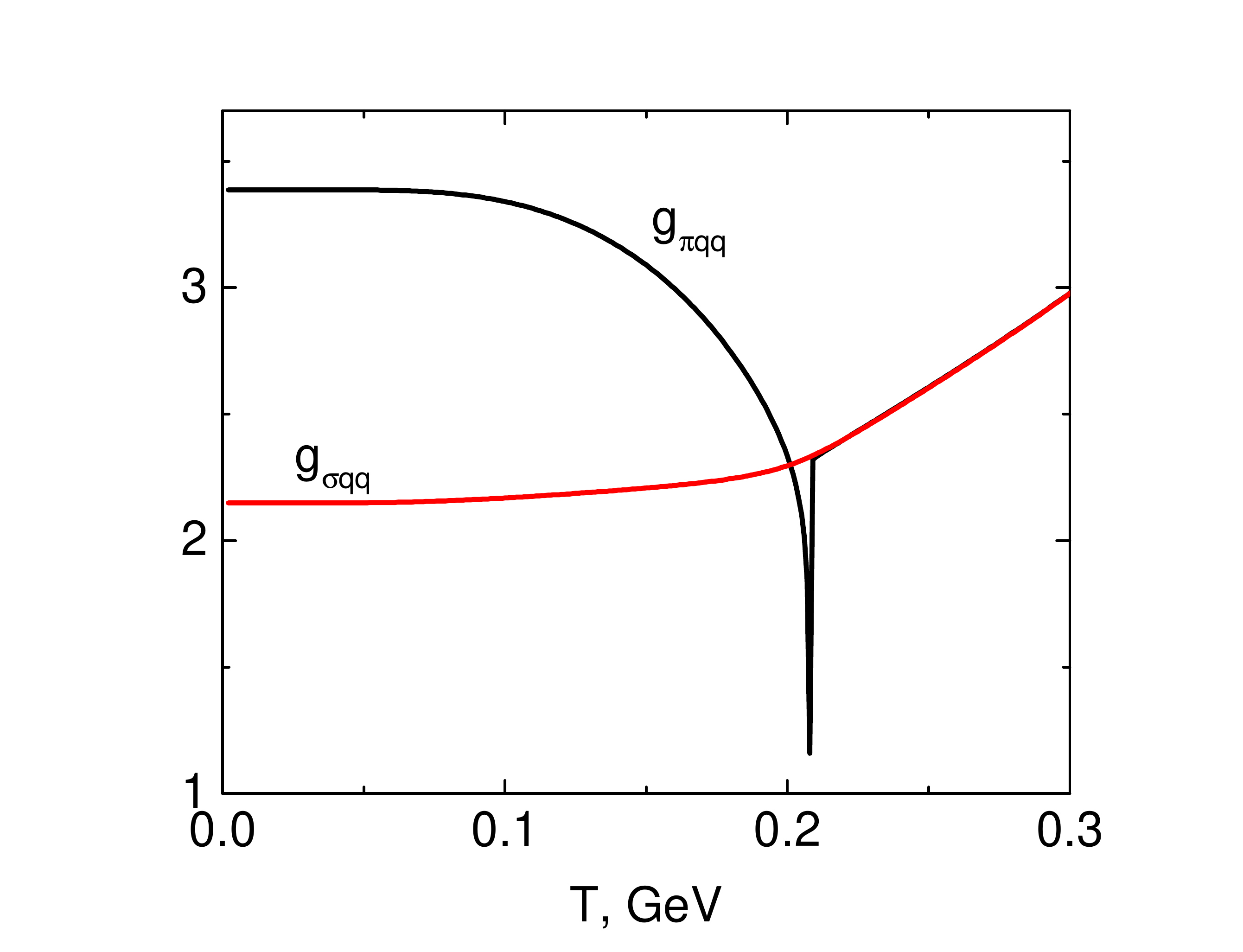}}
   \caption{Left panel: the double quark mass $2m_q(T)$, meson masses $M_\pi(T)$, $M_\sigma(T)$ and their broadening  $M_M\pm\Gamma_M(T)/2$ at $\mu_q = 0$. The vertical dash-dotted lines correspond to the critical temperature and the Mott temperature for pion. Right panel: coupling constants $g_{\pi qq}$,$g_{\sigma qq}$. }
\label{mesmass}
\end{figure}

To describe the mass spectra in the NJL model, a set of parameters is required: the cut-off parameter $\Lambda = 0.639$ GeV, the current quark mass $m_0 = 5.5$ MeV  and the coupling constant $G = 5.227$ GeV$^{-2}$, which are fixed at zero temperature to reproduce some phenomenological values like pion mass, quark condensate and the pion weak decay constant.
Temperature dependence of double quark masses $2m_q$, meson masses $M_\pi\pm\Gamma_\pi/2$ and $M_\sigma\pm\Gamma_\sigma/2$ and coupling constants $g_{\pi qq}$,$g_{\sigma qq}$  as functions of temperature at $\mu_q=0$ are shown in Fig. \ref{mesmass}. The quark mass (or the quark condensate) in the NJL model plays the role of the order parameter for the chiral phase transition, which appears for a given parameter set at $T = 0.192$ GeV \cite{RevModPhys.64.649,Kalinovsky:2015kzf}. At higher temperatures, when $m_q \rightarrow m_0$ the spontaneously broken chiral symmetry is supposed to be partially restored and the mass of the sigma meson tends to the mass of its chiral partner pion. The temperature when meson mass exceeds the mass of the constituents is the Mott temperature. As can be seen in Fig.\ref{mesmass}, for the pion $T^\pi_{\rm{Mott}} = 0.22$ GeV.

\subsection{Scattering amplitude}

The scattering amplitude for the $\pi\pi$ scattering process in the framework of the NJL and PNJL models is described in detail in Refs. \cite{Quack:1994vc,Fu:2009zs}. In the lowest order $1/N_c$ there are two types of Feynman diagrams contributing to the $\pi\pi$-scattering amplitude: four-vertex ``box''-diagrams, described the four-quark interaction  and the meson-exchange diagrams. 
The second type of processes contributing to the $\pi\pi$-scattering amplitude are processes with a meson as an intermediate state.  
As in the $SU(2)$ NJL model only scalar and pseudo-scalar quark-antiquark states are presented, the role of the intermediate state is played only by the scalar $\sigma$-meson, and the triangle vertex corresponds to $\sigma \rightarrow \pi\pi$ decay. The amplitude can be written as  
\begin{equation}
i\mathcal{T}^{\sigma} = i A^{\sigma \pi\pi}D_\sigma(p)i A^{\sigma\pi\pi},
\label{triagTot}
\end{equation} 
where $D_\sigma$ is the meson propagator, 
the triangle amplitude $A^{\sigma\pi\pi}$ defines the coupling strength $g_{\sigma\pi\pi}$= $2 g_{\sigma q q} g_{\pi qq}^2 A^{\sigma\pi\pi}$   \cite{Zhuang:2000tz,Friesen:2011ma}.
The contribution from this diagram plays a role as long as the decay $\sigma \rightarrow \pi\pi$ is possible, in other words as long as the $\sigma$-meson mass exceeds the mass of two pions (in our model this temperature is $T^\sigma_{diss} = 0.189$ GeV).


Using the NJL and PNJL models for description of  the scattering amplitude $|\mathcal{T}|$ at finite temperature is limited by the simplest kinematic $p_1=p_2=p_3=p_4= p$ \cite{Quack:1994vc,Fu:2009zs}. Considering non-trivial kinematics is quite difficult and requires complicated calculations \cite{Rehberg:1995nr,Khvorostukhin:2020foa}. 

Nevertheless, a model truncation is possible. Assuming that "box" diagrams are weakly dependent on the kinematic conditions, the contribution of these diagrams can be replaced by the constant of 4-pion interaction $g_{4\pi}$. Using the pole approximation for the  meson propagator Eq. (\ref{mesonProp}) as
\begin{equation}
D_\sigma(x) \approx \frac{g_{\sigma q q}^2}{M_\sigma^2- x- i\Gamma_\sigma M_\sigma},
\label{propPole}
\end{equation}
and considering the triangle vertex as the coupling constant $g_{\sigma\pi\pi}$, it is possible to show that the scattering amplitudes $T_0, T_1, T_2$ can be written by analogy with the meson exchange model\cite{Cotanch:2002vj}, where quark-exchange diagrams correspond to  contact terms. Then the amplitudes of a particular total isospin $I$ ($I = 0,1,2$) defined as $T_I$   have the form:
\begin{eqnarray}
T_0  &=& 5 g_{4\pi} +  g^2_{\sigma\pi\pi}\left( 3 D_\sigma(s)+D_\sigma(t)+D_\sigma(u) \right),\label{amp_a0} \\
T_1  &=&  g^2_{\sigma\pi\pi}\left( D_\sigma(t) - D_\sigma(u) \right), \label{amp_a1}\\
T_2 & =& 2 g_{4\pi} +  g^2_{\sigma\pi\pi}\left(D_\sigma(t)+D_\sigma(u) \right),
\label{amp_a2}
\end{eqnarray}
where $g_{4\pi}$ is the constant of 4-pion interaction and $D_\sigma(s,t,u)$ is the meson propagator\cite{Bjorken:100769}. 

\begin{figure}[!h]
  \centerline{  \includegraphics[width=0.45\linewidth]{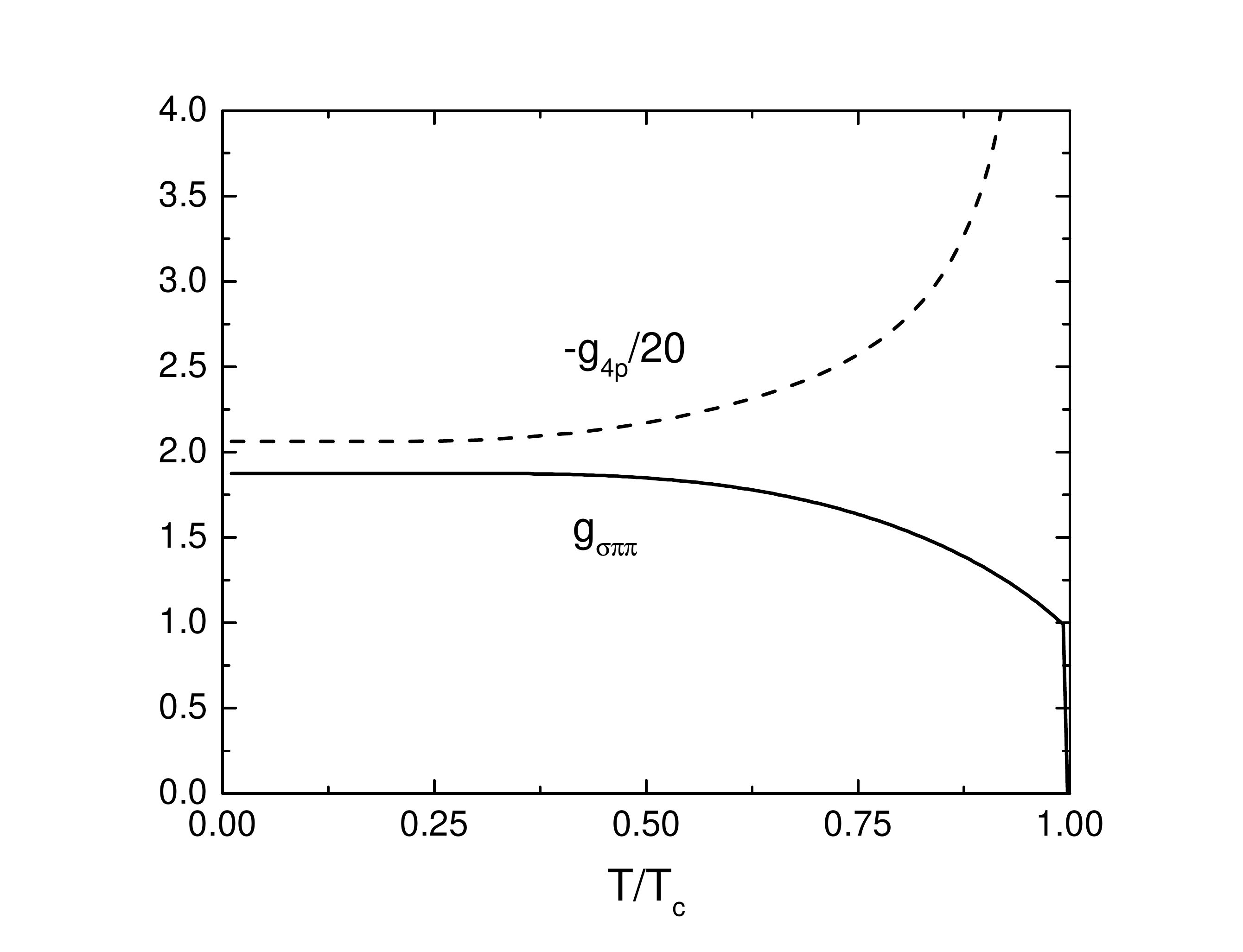}
   \includegraphics[width = 0.44\textwidth]{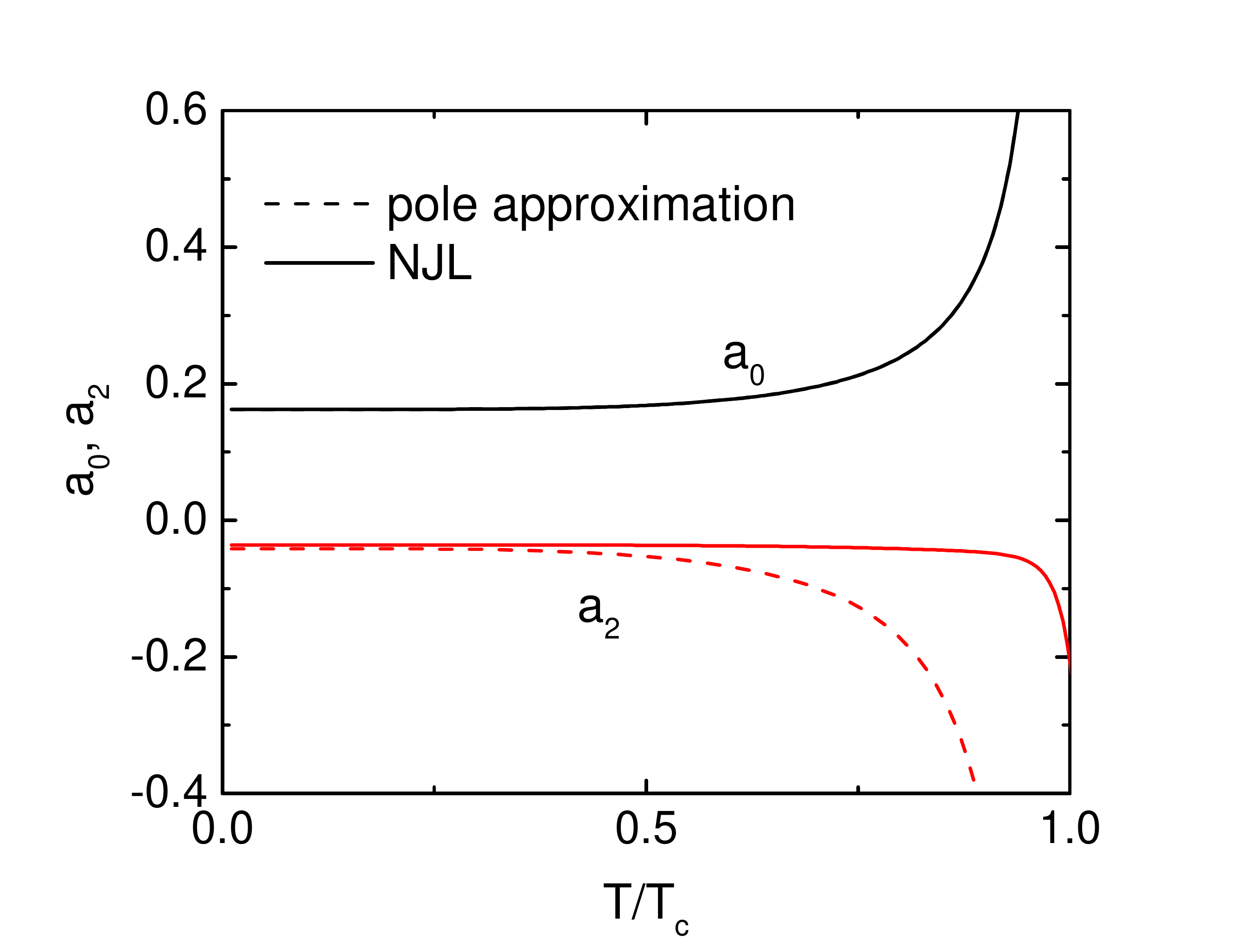}}
  \caption{Left panel: the decay constant $g_{\sigma\pi\pi}$ and  constant $g_{4\pi}$, as functions of temperature. Right panel: scattering lengths $a_0$, $a_2$ calculated with Eqs.(\ref{amp_a0})-(\ref{amp_a2}) (dashed lines) as function of temperatures in comparison with the original NJL results (solid lines).}
\label{couplings}
\end{figure}

The amplitudes $T_0, T_1, T_2$ obtained in the framework of the original NJL model  for the case $t=u=0$ give the scattering lengths $a_0 = T_0/32\pi$, $a_1=0$, $a_2 = T_2/32\pi$. These results are shown in Fig.\ref{couplings}, right panel with solid lines. By inserting the calculated values for $a_0, a_2$, meson masses and coupling constants into Eqs.(\ref{amp_a0}-\ref{amp_a2}), the constant $g_{4\pi}$ can be obtained as a function of temperature. The temperature behaviour of the renewed scattering lengths $a_0$, $a_2$ calculated on the basis of Eqs.(\ref{amp_a0}-\ref{amp_a2})   is shown as dashed lines in the right panel of the Fig.\ref{couplings} in comparison with the NJL results. The results for $a_0$ in the NJL model and the used approximation coincide. The decay constant $g_{\sigma\pi\pi}$ calculated in the NJL model and the constant of four-pion interaction $g_{4\pi}$, normalized to factor $-20$  are shown in the left panel in the Fig. \ref{couplings}. 

Considering a pion-pion collision for a neutral pion, one should take into account the following contributions to the total cross section : $\pi^{0}\pi^{0} \rightarrow \pi^{0}\pi^{0}$, $\pi^{+}\pi^{-}\rightarrow \pi^{0}\pi^{0}$ and $\pi^{\pm}\pi^{0} \rightarrow \pi^{\pm}\pi^{0}$. The physical $\pi\pi$-scattering amplitudes are related to the isospin amplitudes by\cite{Cotanch:2002vj}
\begin{eqnarray}
\mathcal{T}_{\pi^0\pi^0\rightarrow\pi^0\pi^0} &=& \frac{2}{3} T_2+\frac{1}{3} T_0  \\
\mathcal{T}_{\pi^\pm\pi^0\rightarrow\pi^\pm\pi^0} &=&  \frac{1}{2} T_2 + \frac{1}{2}T_1 \\
\mathcal{T}_{\pi^0\pi^0\rightarrow\pi^+\pi^-} &=&  \frac{1}{3} T_2 - \frac{1}{3} T_0,
\end{eqnarray}
from where
\begin{eqnarray}
\mathcal{T}_{\pi^0\pi^0\rightarrow\pi^0\pi^0} &=& 3 g_{4\pi}+ g^2_{\sigma\pi\pi}\left(D_\sigma(s)+D_\sigma(t)+D_\sigma(u)\right) \\
\mathcal{T}_{\pi^\pm\pi^0\rightarrow\pi^\pm\pi^0} &=&  g_{4\pi} +  g^2_{\sigma\pi\pi} D_\sigma(u) \\
\mathcal{T}_{\pi^0\pi^0\rightarrow\pi^+\pi^-} &=&  g_{4\pi} -  g^2_{\sigma\pi\pi} D_\sigma(s).
\end{eqnarray}


\label{sec:preparation}
\section{Pion damping width and pion spectral function}

The description of the resonance properties at finite temperature and density was carried out by Kadanoff and Baym \cite{KadanoffBaym}. The width $\Gamma$ of a particular state (or lifetime $\tau$) can be calculated by the equation 
\begin{equation}
\Gamma(p) = \tau^{-1}(p) =\Sigma^{>}(p)\pm\Sigma^{<}(p),
\end{equation}
where the sign ''$-$"  is used for bosons and ''$+$'' for fermions. The lifetime of particular state in a dense medium depends on the probability of its decay into other states and the probability of  inverse processes, which can restore the decayed state and thereby can prolong the lifetime of the state. The ''+'' for fermions appears due to the Pauli principle, which prevents the inverse processes of scattering into the existing state. For bosons, there appears a competition between the processes which tends to increase the number of existing states (described by the function $\Sigma^<$ ) and the processes which tends to decrease them  ($\Sigma^>$). The functions $\Sigma^<$, $\Sigma^>$ have the form  \cite{KadanoffBaym}
\begin{eqnarray}
 \Sigma^{<}(p) &=& \int_{p_1}\int_{p_3}\int_{p_4}(2\pi)^4\delta_{p_1,p_2;p_3,p_4}|T|^2G^>(p_1)G^<(p_3)G^<(p_4), \label{func_sigma1}\\  
 \Sigma^{>}(p)&=& \int_{p_1}\int_{p_3}\int_{p_4}(2\pi)^4\delta_{p_1,p_2;p_3,p_4}|T|^2G^<(p_1)G^>(p_3)G^>(p_4),
 \label{func_sigma2}
\end{eqnarray} 
where the functions $G_i^>=[1+n_i(\omega)]A_i(p^2)$, $G_i^<=n_i(\omega)A_i(p^2)$ define the average density of particles with momentum $\bar{p}$ and energy $\omega$, and $n_i = (\rm{exp} (\beta\omega)- 1)^{-1}$ is the particles occupation number for bosons. Some notations were introduced for brevity: $\int_{p_i} = \int\frac{dp_i}{(2\pi)^4}$, $\delta_{p_1,p_2;p_3,p_4} = \delta(p_1+p_2-p_3-p_4)$, with $p_2=p$, as $\Gamma$ is calculated in the rest framework for particle number 2. 

After integration over zero-components of momenta, integrals (\ref{func_sigma1}), (\ref{func_sigma2}) take a form
\begin{eqnarray}
&& \Sigma^{<(>)}(m_2,\vec{0}) =  \\
 && = \frac{1}{64\pi^4}\int d \Omega  \frac{d \vec{p_1}}{2 E_1}\int_{-1}^1  d (\cos\alpha)\frac{|\vec{p_3^*}|^2\cdot|\mathcal{T}|^2}{|\vec{p_3^*}|(E_1+m_2)-|\vec{p_1}||\vec{p_3^*}| \rm{cos}\alpha}F^{<(>)} \nonumber \label{func_sigma1_int},
\end{eqnarray}
where $d\Omega= ds_1 A(s_1) ds_3 A(s_3) ds_4  A(s_4)$, the factors $F^>= n_1 (n_3+1)(n_4+1)$ and $F^<=(n_1+1)n_3 n_4$ are introduced for brevity, the momentum $\vec{p_3^*}$ is defined as:
\begin{eqnarray}
|\vec{p_3^*}|_{1,2} =
\frac{|\vec{p_1}|a b \pm \sqrt{|\vec{p_1}|^2 a^2 b^2  +((\sqrt{s_2})+E_1 )^2-|\vec{p_1} |^2 b^2)(a^2-4s_3 (\sqrt{s_2}+E_1 )^2 ) }}{2((\sqrt{s_2}+E_1 )^2 -|\vec{p_1} |^2 b^2)}.   \nonumber
\label{eq_p3}
\end{eqnarray}
with $a=(\sqrt{s_2 }+s_1 )^2+s_3-s_4+s_1- E_1^2$, $s_i = m_i^2$ and $b = \cos\alpha$. The spectral function A ($s_i$) is chosen in the Breit–Wigner form 
\begin{equation}
A(s) =\alpha_s\frac{M\Gamma}{(s-M^2)^2+M^2\Gamma^2},
\label{Breit_Wigner}
\end{equation}
where $M$ is the meson pole mass, $\Gamma$ is the corresponding meson width, and $\alpha_s$ is a normalization factor, $\alpha_s=2$.

\begin{figure}[!h]
  \centerline{  
   \includegraphics[width = 0.5\textwidth]{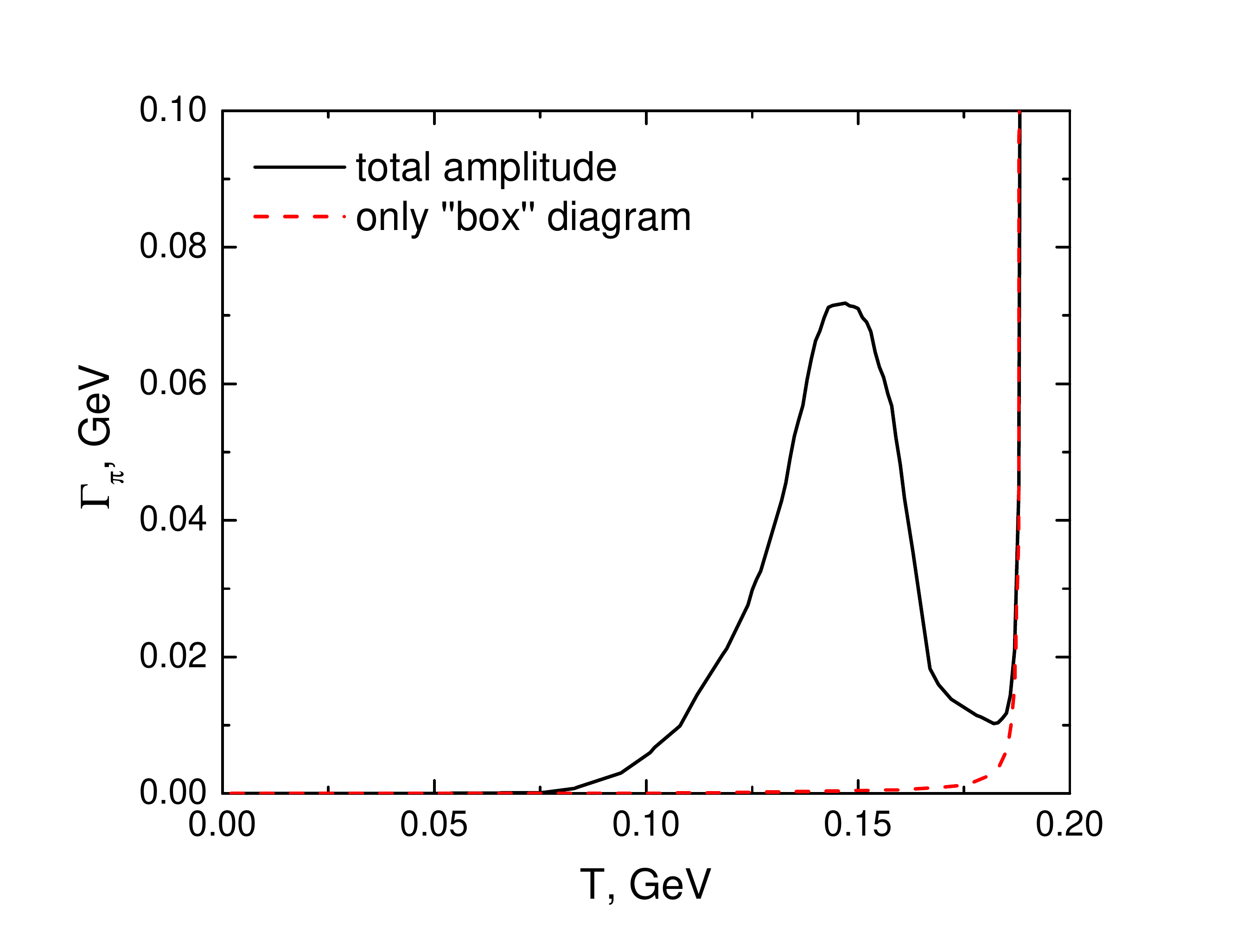} 
   \includegraphics[width = 0.5\textwidth]{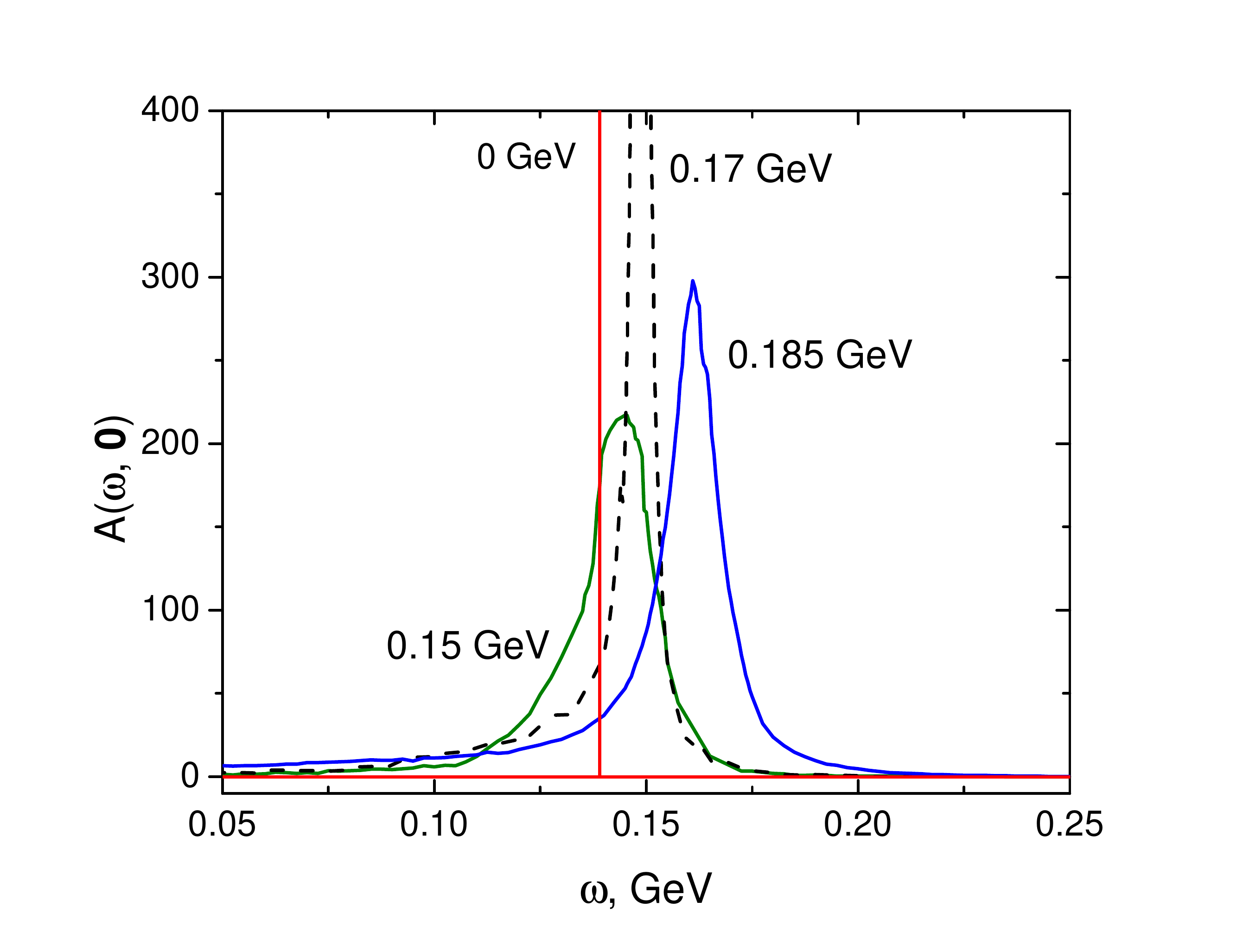}}
  \caption{Left panel: the pion width as functions of temperature. The red dotted line corresponds to the case when amplitude contains only ''box''-diagrams.  Right panel: the pion spectral function $A(\omega,\bar{0})$ at different temperatures. }
\label{Gamma_Aw}
\end{figure}

It is clearly seen in the left panel of Fig.\ref{Gamma_Aw}, that the pion state broadens in a hot pion gas after T$\sim 0.1$ GeV and reaches the maximal value $\Gamma\sim 0.075$ GeV at $T\sim 0.145$ GeV, then the curve turns down.  This behaviour can be caused by  the properties of the scalar $\sigma$- resonance, which plays a significant role in the pion-pion scattering.  As was said before, the calculations in our model are limited by $T\sim 0.19$ GeV, where the chiral phase transition takes place and the $\sigma\rightarrow\pi\pi$ decay is limited by $T\sim 0.189$ GeV, after which only the "box"-diagram participates in the scattering amplitude. The pion width increases again near the critical temperature, as pion at high temperature tends to melt. The results obtained for the amplitude with taking into account only "box"-diagrams are shown in  Fig. \ref{Gamma_Aw} with dashed line.

We would like to pay some attention  to the spectral function $A_i(\omega,\bar{p})$, which defines the possible energy spectrum $\omega$ for a particle with momentum $\bar{p}$. For a non-interacting gas the spectral function can be used as a $\delta$-function, but for a more general case it should be considered as \cite{KadanoffBaym} 
\begin{equation}
A(\omega, {\bar{p}}) = \frac{\Gamma(\omega, \bar{p})}{(\omega-E(\bar{p})-{\rm Re}\Sigma_c(\omega,\bar{p}))^2+(\frac{\Gamma(\omega, \bar{p})}{2})^2}.
\end{equation} 
To find the spectral function $A(\omega,\bar{p})$, a self-consistent set of functional equations has to be solved. The set contains equations for  $\Gamma(\omega,\bar{p})$, $A(\omega,\bar{p})$ and self-energy $\Sigma_c(\omega,\bar{p})$
\begin{equation}
{\rm Re}\Sigma_c(\omega, {\bar{p}}) =\mathcal{P}\int\frac{d\omega'}{2\pi}\frac{\Gamma(\omega',\bar{p})}{\omega-\omega'},
\end{equation} 
 which again depends on the function of $\Gamma(p)$. All functions depend on energy and momentum,  and  the set of functional equations is complicated to solve.
 
For simplicity, the set of equations can be solved by the iteration method. The first step starts from $A(\omega, {\bar{p}})$ in the approximation of small $\Gamma$, for which the Breit-Wigner form (Eq. (\ref{Breit_Wigner})) is used. For the second step, the obtained $\Gamma$ and $A$ are put back into Eqs.(\ref{func_sigma1}, \ref{func_sigma2}) to get the second approximation for $\Gamma$ and obtain the second iteration for the spectral function $A(\omega, {\bar{p}})$. 

The spectral functions at several temperatures are shown in Fig.\ref{Gamma_Aw}, right panel. As at T=0 the pion is a bound state, its spectral function is almost $\delta$-function, which is located at the pole corresponding to the pion mass. With increasing temperature, the width increases and  the pion becomes a resonant state. As at $T>0.15$ GeV the width goes down again, the spectral function at $T\sim 0.17$ GeV
is higher and narrower than at $T=0.15$ and $T = 0.188$ GeV. As we mentioned above, we are limited in our approximation of elastic scattering and are  not able to calculate the spectral function at temperatures higher than the critical temperature. For such research the approximation of the scattering phase shift is more correct, as this approximation controls both the bound state of the meson and the quark background\cite{Xia:2014bla}.

\section{Conclusion and outlooks}  

The in-medium modification of hadron properties at high temperature and density can affect the observables and should be taken into account in the analysis of experimental data. One of these properties is the finite meson width, that can appear in dense matter. For light particles the increase in temperature leads to an increase of density, and collisions of particles occur at a much higher rate. This leads to an increase  in the probability of both direct processes, tending to decrease the lifetime of a particular state (or increase its width) and the inverse processes, tending to increase it. The resulting width is related to the damping - the process of returning  a disturbed many-particle system to an equilibrium state.

In this work in a self-consistent approach in the framework ofthe NJL model, the pion damping width and the pion spectral functions are considered. It is shown in Fig. \ref{Gamma_Aw}, the pion width at $T<0.17$ GeV depends on the inclusion of the $\sigma$-meson, namely allowance of the decay $\sigma\rightarrow\pi\pi$ and the induced decay $\pi+\pi\rightarrow\sigma$. The dashed line shows that the absence of the $\sigma$-exchange diagram in the total scattering amplitude does not demonstrate the broadening of the pion width in this area. At $T\rightarrow T_c$, the system tends to the deconfinement phase and the quark-antiquark  pion structure becomes more significant: the pion is considered rather a resonant state and the pion width rises up. We find that at critical temperatures $T>0.19$ GeV our approach ceases to be valid, and for a more detailed study of the width and spectral function one should use the quark-antiqurk phase shift formalism developed in Refs.\cite{Xia:2014bla,Blaschke:2019col,Blaschke:2013zaa}.

As a final step, we try to evaluate the effect of the $\rho-$meson on the damping pion width. It is interesting, as  a significant effect  of the including of the finite pion width  in the observables of a $\rho-$meson is discussed in literature \cite{vanHees:2000bp}. An evaluation of the effect of the vector channel on the $\pi\pi$ scattering amplitude in the framework of the NJL model has a lot of difficulties as the including of the $\rho$-meson  strongly depend on the parameters, and on the quark mass affecting the $\rho-$meson to be a bound state \cite{He:1997gn,Jaminon:2002dx}. For simplicity, we use the mass equation obtained in the low-momentum expansion obtained in  Ref.\cite{Ebert:1992ag}, supposing the width $\Gamma_{\rho qq} = 0$ GeV.
\begin{figure}[!h]
  \centerline{  
   \includegraphics[width = 0.6\textwidth]{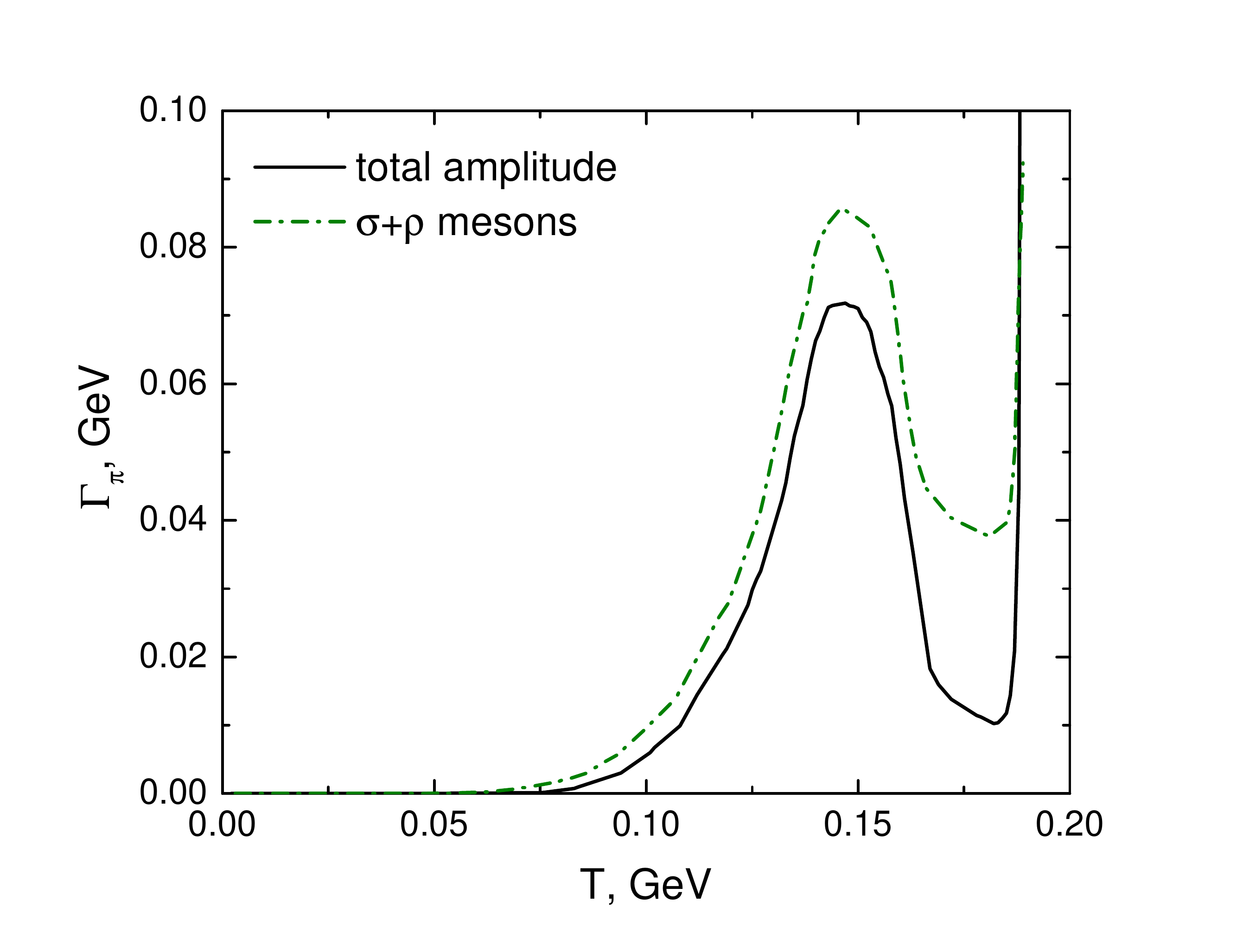}}
  \caption{The pion width as functions of temperature with taking into account the $\rho$-meson channel (green dot-dashed line) in comparison with scalar channel (black solid line).}
\label{Gamma_rho}
\end{figure}

The partial amplitudes $T_0, T_1, T_2$ in Eqs.(\ref{amp_a0}-\ref{amp_a2}) should be extended by the $\rho-$channel additional terms \cite{Cotanch:2002vj}:
\begin{eqnarray}
T_0^\rho  &=&   2 g^2_{\rho\pi\pi}\left( (u-s) D_\rho(t)+ (t-s) D_\rho(u) \right),\label{amp_a0_rho} \\
T_1^\rho  &=&  g^2_{\rho\pi\pi}\left(2 (u-t) D_\rho(s) + (u-s) D_\rho(t) +(s-t) D_\rho(u) \right), \label{amp_a1_rho}\\
T_2^\rho & =&   g^2_{\rho\pi\pi}\left((s-u)D_\rho(t)+(s-t)D_\rho(u) \right),
\label{amp_a2_rho}
\end{eqnarray}
with the meson propagator $D_\rho(x) \approx g_{\rho q q}^2(M_\rho^2- x)^{-1}$ and $M_\rho$, $g_{\rho q q}$ calculated in accordance with Ref. \cite{Ebert:1992ag}, $g_{\rho\pi\pi}= 5.14$. The result of a such simple iteration for including the vector channel to the pion damping width is presented in Fig. \ref{Gamma_rho} with green dot-dashed line. As it was shown in Ref.\cite{Jaminon:2002dx}, the inclusion of the $\rho-$channel in the $\pi\pi$-scattering amplitude in the low-momentum approach has non-vanishing contribution only to the $a_1^1$ scattering length, which is defined by the $T_1$ isospin amplitude and does not contribute to the $a_0^0$ and $a_0^2$ ($T_0, T_2$ amplitudes). So, in this simplest approach we should not expect a strong effect to the pion damping width from the $\rho-$meson. And a more detailed description of the $\rho$-meson and $\pi-a_1$ mixing can give a more significant effect to the behaviour of the pion damping width in the hot pion gas.

\end{document}